\documentclass{aa}  

\usepackage{graphicx}
\usepackage{txfonts}
\begin{document}

   \title{The environment of the wind-wind collision region of $\eta$ Carinae}

   \author{C. Panagiotou
          \and
          R. Walter
          }

   \institute{
              Astronomy Department, University of Geneva, Chemin d’Ecogia 16, 1290 Versoix, Switzerland\\
              \email{Christos.Panagiotou@unige.ch}         
             }

   \date{Received ***; accepted ***}

 
  \abstract
   {$\eta$ Carinae is a colliding wind binary hosting two of the most massive stars and featuring the strongest wind collision mechanical luminosity. The wind collision region of this system is detected in X-rays and $\gamma$-rays and offers a unique laboratory for the study of particle acceleration and wind magneto-hydrodynamics.}
   {Our main goal is to use X-ray observations of $\eta$ Carinae around periastron to constrain the wind collision zone geometry and understand the reasons for its variability.}
   {We analysed 10 Nuclear Spectroscopic Telescope Array (NuSTAR) observations, which were obtained around the 2014 periastron. The NuSTAR array monitored the source from 3 to 30 keV, which allowed us to grasp the continuum and absorption parameters with very good accuracy. We were able to identify several physical components and probe their variability.}
   {The X-ray flux varied in a similar way as observed during previous periastrons and largely as expected if generated in the wind collision region. The flux detected within $\sim 10$ days of periastron is lower than expected, suggesting a partial disruption of the central region of the wind collision zone. The Fe K$\alpha$ line is likely broadened by the electrons heated along the complex shock fronts. The variability of its equivalent width indicates that the fluorescence region has a complex geometry and that the source obscuration varies quickly with the line of sight.}
  {}

   \keywords{Stars: individual: $\eta$ Carinae -- Stars: winds, outflows -- X-rays: binaries}

   \maketitle
%

\section{Introduction}

$\eta$ Carinae is a highly eccentric (e $\sim$ 0.9) binary stellar system with an orbital period of 5.54 years. This binary system hosts one of the most massive stars, which has an estimated initial mass above 100$\hspace{2pt}\text{M}_\odot$ \citep{2001ApJ...553..837H}, and is classified as a luminous blue variable. The companion star has also a high mass of $\sim30 \hspace{2pt} \text{M}_\odot$. This companion has not been directly observed, but it is thought to be an O supergiant or a Wolf-Rayet star. 

The binary system is located inside a bipolar nebula, the so-called Homunculus Nebula (HN; hereafter), which was formed by the Great Eruption in the 1840s. During this period the system underwent a massive outburst with a total mass loss rate estimated as $\sim1 \hspace{2pt}\text{M}_\odot$/yr \citep{2003AJ....125.1458S}. The physical origin of this outburst is still unclear. 

The two stars eject winds with a significant mass loss rate. The wind of the primary star is dense and slow and has a mass loss rate of $\sim8.5 \times 10^{-4}\hspace{2pt}\text{M}_\odot \text{yr}^{-1}$ and a terminal velocity of $\sim420 \hspace{2pt}\text{km/s}$ \citep{2012MNRAS.423.1623G}. The secondary wind is much faster, has a terminal velocity of $\sim3000 \hspace{2pt}\text{km/s}$, and corresponds to a mass loss rate of $\sim10^{-5}\hspace{2pt}\text{M}_\odot \text{yr}^{-1}$ \citep{2002A&A...383..636P}. The two winds collide and form the so-called wind-wind collision (WWC, hereafter) region. The shock heats the surrounding plasma, which then emits X-rays.

Beyond the WWC region, two more components contribute to the X-ray emission. The first component corresponds to the reflection of the central emission by the HN. In an analysis of Chandra observations, \cite{2004ApJ...613..381C} found that the HN spectrum is well reproduced by a two-temperature plasma emission. The second component is the constant central emission (CCE, hereafter) discussed by \cite{2007ApJ...663..522H}. Recently, \cite{2016MNRAS.458.2275R} showed that the CCE is produced by the collision of the secondary wind with part of the primary wind ejected in the previous orbit and by the leading arm of the current WWC shock. These two sources, CCE and HN, are relatively faint and mainly observed during the X-ray minimum close to periastron. 

Because of the high orbital eccentricity, the stellar separation varies by a factor of 20. As a result, the WWC emission is also highly variable, since it is proportional to the inverse of this distance \citep{1992ApJ...386..265S}. The X-ray variability has been studied for over two decades, providing information for the last four orbital cycles. In every cycle, the emission was found to vary in a similar way \citep{2017ApJ...838...45C}, reaching a flux maximum several days prior to periastron and then steeply dropping to a minimum state before returning to an intermediate level. The duration of the minimum state is of the order of a few weeks and differs in each cycle. The origin of the minimum remains unclear. A significant increase of the source extinction or a partial disappearance of the WWC region have been suggested. 

\cite{2016ApJ...817...23H} studied two observations of $\eta$ Carinae performed simultaneously by NuSTAR and XMM-Newton just before the 2014 periastron ($\phi$= 2.9721 and 2.9979). These authors found that the WWC emission can be described by a highly absorbed two-temperature plasma emission and suggested that the X-ray minimum is probably due to an increased absorption of the source in our line of sight. These authors also discussed the non-detection of the high-energy power-law emission of $\eta$ Carinae by NuSTAR in these phases.

In this work, we analysed NuSTAR \citep{2013ApJ...770..103H} observations of $\eta$ Carinae. The NuSTAR array provides the first opportunity to study simultaneously soft (<10 keV) and hard (>10 keV) X-ray variability during periastron and also to investigate the physical origin of the minimum. Section 2 describes the data reduction. The results are discussed in Section 3 and summarised in Section 4.

\section{Data reduction}

\subsection{NuSTAR data}

The NuSTAR array monitored $\eta$ Carinae before and after the 2014 periastron.
In total, 10 observations were performed from March 31, 2014 to July 16, 2015 with an exposure ranging from 23.6 to 81.4 ks. The details of each observation are listed in Table \ref{tab:obs_info}. The phase was estimated as $\phi = (\text{MJD} - 50799.292)/2024$ \citep{2005AJ....129.2018C}. The NuSTAR array has two detectors, namely the FPMA and FPMB, which are both sensitive to photons with energy from 3 to 78 keV. In this work we only used the data gathered by FPMA. 

\begin{table}
        \centering
        \caption{Date (middle time) and exposure of each observation. The third column lists the corresponding orbital phase.}
        \label{tab:obs_info}
        \begin{tabular}{cccc} 
                \hline
                \noalign{\smallskip}
                {Obs id}     &   MJD       & $\phi$   &  Exposure (ks) \\
                \noalign{\smallskip}
                \hline
                \noalign{\smallskip}
                30002010002  &  56747.587  &  2.9389  &  28.8    \\
                30002010003  &  56748.429  &  2.9393  &  49.8    \\
                30002010005  &  56804.245  &  2.9669  &  79.4    \\
                30002040002  &  56814.739  &  2.9721  &  32.9    \\
          30002040004  &  56867.038  &  2.9979  &  61.4    \\ 
                30002010007  &  56880.584  &  3.0046  &  31.0    \\ 
                30002010008  &  56881.608  &  3.0051  &  57.0    \\
                30002010010  &  56889.268  &  3.0089  &  54.5    \\
                30002010012  &  56926.857  &  3.0274  &  81.4    \\
                30101005002  &  57219.303  &  3.1719  &  23.6    \\
                \noalign{\smallskip}
                \hline
        \end{tabular}
\end{table}

\begin{figure}
  \centering
  \includegraphics[width=140pt]{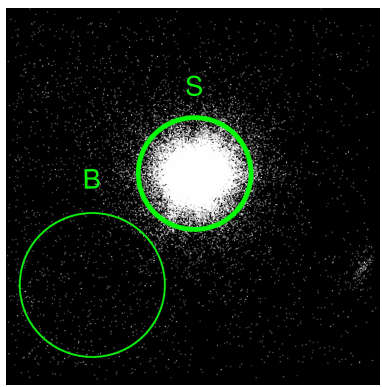}
  \caption[10]{Full-band NuSTAR image of the first observation. The ’S’ and ’B’ circles correspond to the source and background extraction regions, respectively. The image is in logarithmic scale and covers the total field of view of the instrument.}
  \label{fig:img_10002}
\end{figure}

In order to produce high-level data (i.e. light curves, spectra, and response files) we used the NuSTAR Data Analysis Software (NuSTARDAS) version 1.6.0. We visually examined the sky images to decide on the source and background extraction regions. The full-band image of the first observation is shown in Fig. \ref{fig:img_10002}. Its size corresponds to the entire NuSTAR field of view (12.2$\times$12.2 arcmin). The green circles are the source ($\sim$2.07 arcmin) and background ($\sim$2.66 arcmin) extraction regions. 

\begin{figure*}
  \centering
  \includegraphics[width=300pt,clip]{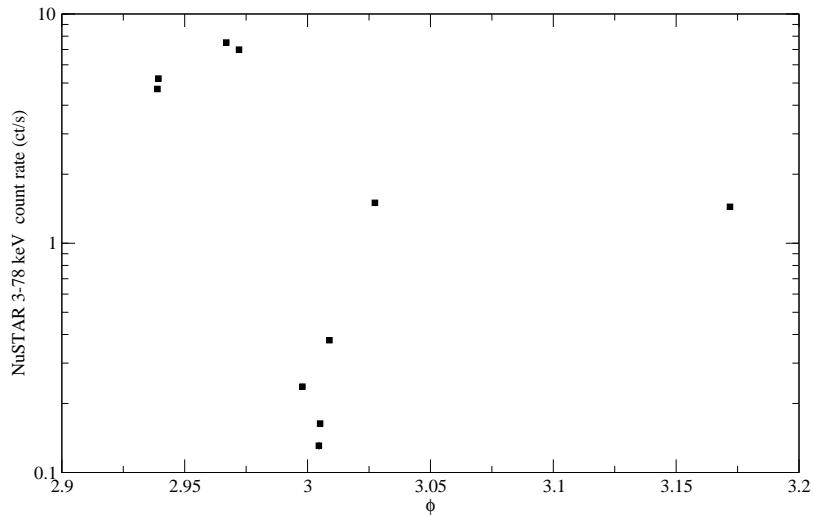}
  \caption[10]{ $\eta$ Carinae light curve from NuSTAR. The horizontal axis is the orbital phase. The count rate uncertainties are too small to be distinguished.}
  \vspace{0.5cm}
  \label{fig:lc_all}
\end{figure*}

With the exception of the third observation, the source did not present significant variability during any of the observations. The overall light curve was highly variable, as expected. Figure \ref{fig:lc_all} shows the total light curve of $\eta$ Carinae. Each point corresponds to an individual observation. The flux increased to a maximum and decreased steeply a few days before periastron. The minimum value of the emission was observed nine days after periastron, while about two months later the flux returned to an intermediate level.

\subsection{BAT and ISGRI data}

In order to investigate the hard X-ray emission of $\eta$ Carinae, we also analysed data obtained with the Swift/BAT \citep{0067-0049-209-1-14} and INTEGRAL/IBIS/ISGRI \citep{2003A&A...411L.141L} instruments. Their wide field of view allows the accumulation of large exposure time for any sky position (in the Galactic plane for ISGRI).

The Swift/BAT reduction pipeline is described in \cite{2010ApJS..186..378T} and \cite{2013ApJS..207...19B}. Our pipeline is based on the BAT analysis software HEASOFT v 6.13. A first analysis was performed to derive background detector images. We created sky images (task \texttt{batsurvey}) in the eight standard energy bands (in keV: 14 - 20, 20 - 24, 24 - 35, 35 - 50, 50 - 75, 75 - 100, 100 - 150, and 150 - 195) using an input catalogue of 86 bright sources that have the potential to be detected in a single pointing. The detector images were then cleaned by removing the contribution of all detected sources (task \texttt{batclean}) and averaged to obtain one background image per day. The variability of the background detector images was smoothed pixel-by-pixel fitting the daily background values with different functions (e.g. spline and polynomial). A polynomial model with an order equal to the number of months in the data set adequately represents the background variations. The BAT image analysis was then run again using these smoothed averaged background maps. The new sky images were stored in an all-sky pixel database by properly projecting the data on a fixed grid of sky pixel, preserving fluxes; the angular size of the BAT pixels varies in the field of view. This database can be used to build local images and spectra or light curves for any sky position.
The result of our processing was compared to the standard results presented by the Swift team (light curves and spectra of bright sources from the Swift/BAT 70-months survey catalogue\footnote{\texttt{http://swift.gsfc.nasa.gov/results/bs70mon/}}) and a very good agreement was found.

The ISGRI analysis is described in \cite{2003A&A...411L.223G}. We used the OSA (v. 9) analysis products available through the HEAVENS database \citep{2010int..workE.162W}.

The final steps of the analysis are identical for both ISGRI and BAT. For each time bin a weighted mosaic of the selected data is first produced and the source flux is extracted assuming fixed source position and shape of the point spread function. We built a weighted mosaic of the selected data in the standard energy bands to obtain average source spectra.

\begin{figure}
  \includegraphics[width=\linewidth,clip]{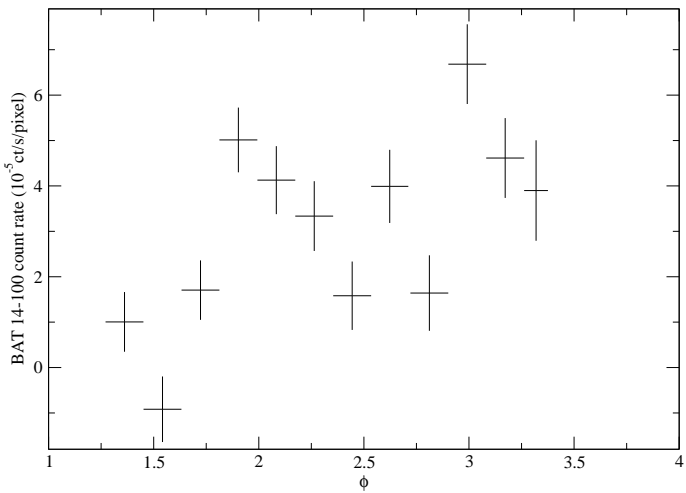}
  \caption{$\eta$ Carinae 14-100 keV light curve  as measured by Swift/BAT. The horizontal axis is the orbital phase.}
  \label{fig:lc_bat_isgri}
\end{figure}

The BAT (14--100 keV) light curve of $\eta$ Carinae shows local maxima close to periastron (see Fig. \ref{fig:lc_bat_isgri}). These variations are dominated by the $<30$ keV thermal emission. The emission above 30 keV, as depicted in Fig. \ref{fig:bat_isgri_spec}, is too weak to allow for a variability analysis.

\section{Results}

\subsection{Spectral fitting}

In the following, we fitted the NuSTAR spectra with an appropriate physical model. As mentioned in the introduction, the X-ray emission of $\eta$ Carinae is the combination of three different components: 1) the  CCE, 2) the reflection of the HN, and 3) the emission produced by the WWC region. Our first goal was to estimate the spectrum of the CCE and HN emission,  expected to be primarily detected just after periastron, when the WWC emission is at its lowest level. We, thus, assumed that the continuum detected during the sixth observation (TJD 16880.584) is mostly the result of these two emission components and fitted the corresponding spectrum by an absorbed two-temperature thermal collisional plasma model.  We also added two narrow Gaussian profiles at energies 6.4 and 7.06 keV to account for the Fe K$\alpha$ and K$\beta$ lines, respectively, and an absorption Fe edge at 7.1keV. The HN and CCE emission are known to be more complex \citep{2004ApJ...613..381C,2016MNRAS.458.2275R}, and thus our model is a simplification limited by the quality of the data.

We used the XSPEC software \citep{1996ASPC..101...17A} version 12.9.0 for the spectral fitting. The applied model in XSPEC terminology is \textit{edge*wabs*(apec+apec+gauss+gauss)}. The model fits well the observed spectrum in terms of $\chi^2$ statistics, i.e. $\chi^2 = 93$ with 69 degrees of freedom (dof, hereafter), which corresponds to a null hypothesis probability of 2.9\%. The best fit results in a temperature of $\text{T}_{\text{c,0}} = 0.55^{+0.33}_{-0.28}\hspace{2pt} \text{keV}$ for cooler plasma and $\text{T}_{\text{h,0}} = 6.5 \pm 1.4 \hspace{2pt} \text{keV}$ for hotter plasma. All the errors quoted in this work correspond to a one parameter, 1$\sigma$ (68\%) confidence interval.

While our model is only a proxy of the true radiation processes, the resulted temperatures are similar to those found in more detailed studies \citep[see for example][]{2015EAS....71...37H,2004ApJ...613..381C} 

We assumed that the CCE and HN emission are constant throughout the whole monitoring period and we used the best-fit model found above to account for those in all spectra.

Having estimated the CCE and HN contributions, we fitted the spectra using the XSPEC model \textit{edge*[(CCE+HN)+wabs*(gauss+gauss+apec+nei+powerlaw)]}. The term \textit{(CCE+HN)} indicates the thermal component of the model used to fit the CCE and HN emission. Its parameters are fixed to the values found previously. The two Gaussians model the Fe K$\alpha$ and K$\beta$ lines again with the normalisation parameter of the former line left free to vary. Following \cite{2016ApJ...817...23H} we fixed the normalisation of the Fe K$\beta$ line as 12\% of that of the Fe K$\alpha$ line \citep{2014ApJ...785L..27Y}. The energy of the absorption edge is fixed to 7.1 keV, while the absorption depth is left free to be minimised. The \textit{apec+nei} component models the WWC emission. This component simulates the emission by a two-temperature collisional plasma emission with the one component in a non-equilibrium state. Finally, the \textit{powerlaw} model accounts for the high-energy power-law emission, detected in $\eta$ Carinae by INTEGRAL and Suzaku. 

For simplicity, we assumed that the abundances of the two plasmas are equal and remain constant through the observation period. We also forced the following parameters to be the same in every observation: 1) the widths of the Fe lines, 2) the ionisation timescale of the \textit{nei} model, 3) the normalisation of the power law, and 4) the temperatures of the different plasmas. We linked those parameters because the quality of the data is not high enough to allow for a direct determination of these values during the fitting procedure. The photon index of the power law was kept frozen to the value 1.8, which is the best-fit value determined by \cite{2010A&A...524A..59L}. 

We found that the best-fit values for the temperatures are $T_{c} = 3.43_{-0.03}^{+0.08}\hspace{2pt} \text{keV}$ and $T_{h} = 5.58 \pm 0.03\hspace{2pt} \text{keV}$ for the cooler and hotter component, respectively, while the cooler plasma corresponds to the component that is in equilibrium. The abundance of the plasma region was estimated to be $0.428 \pm 0.06$ in units of solar abundance and the Fe line width was found to be unexpectedly high with $\sigma = 0.217 \pm 0.006\hspace{2pt} \text{keV}$. Interestingly, the best-fit 1$\sigma$ confidence interval for the average power-law normalisation at 1 keV was [$0,7.3\times10^{-6}$] photons/keV/cm$^2$/s. This upper limit is significantly lower than the value estimated by \cite{2010A&A...524A..59L}, indicating that the power-law emission was not detected by NuSTAR.

\begin{figure}
  \centering
  \vspace{0.15cm}
  \includegraphics[width=\linewidth,height=170pt,clip]{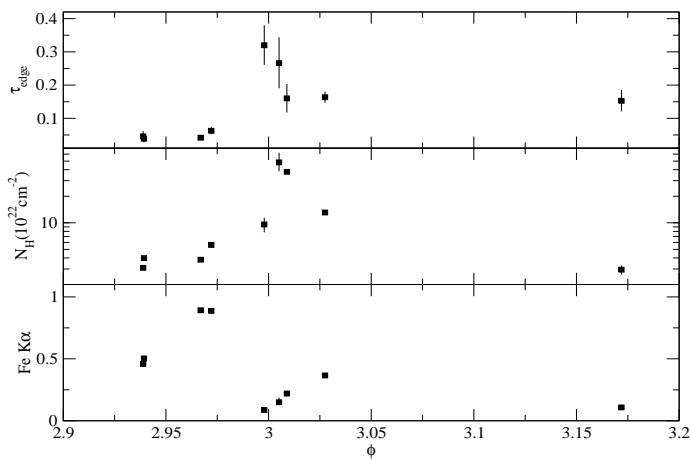}
  \caption[10]{Best-fit values of the edge absorption depth (upper panel), hydrogen column density (middle panel), and normalisation of the Fe K$\alpha$ line (bottom panel) in units of $10^{-3}$ photon/s/cm$^{2}$.}
  \label{fig:best_fit_1}
\end{figure}

The best-fit values of the model parameters left free to be minimised are listed in Table \ref{tab:best-fit} and plotted in Fig. \ref{fig:best_fit_1}. Both the depth of the edge and the absorption column density are proportional to the amount of matter surrounding the WWC source and we expected them to have a similar variability, as observed. It is also reasonable that their values during periastron are higher than in the rest of the observations. Moreover, the iron line normalisation presents a time variability similar to that of the observed flux (see figures \ref{fig:best_fit_1} and \ref{fig:lc_all}). This point is discussed further in section \ref{sec:Fe_line}.

Our results are similar to those found in previous studies. In particular, the best-fit values for the two temperatures and the elements abundance lie reasonably close to the values reported by \cite{2016ApJ...817...23H}. Although the differences between the models do not allow for an accurate comparison of the absorption column densities, our best-fit $N_H$ values do not differ from earlier estimates \citep[e.g.][]{2007ApJ...663..522H,2014ApJ...795..119H}. Also, the absorption variability is qualitatively in agreement with the predictions from hydrodynamics simulations \citep{2011ApJ...726..105P}

In general, the quality of the fit was moderate with $\chi^2 / dof = 3214 / 2630$. We visually examined the residuals; however, no specific trend was prominent. Taking into account the complexity of the source and simplicity of our model, we considered the fit to be representative.

\begin{table*}[h]
        \centering
        \caption{Best-fit values of the variable model parameters. The first column lists the orbital phase. The second, third, and fourth columns list the best-fit values for the edge absorption depth, the hydrogen column density and the normalisation of the Fe K$\alpha$ line, respectively. The normalisations for the models used to fit the WWC emission (see text for details) are listed in the two columns before the last. The last column lists the equivalent width of the Fe K$\alpha$ line. The errors correspond to 1$\sigma$ confidence interval. }
        \label{tab:best-fit}
        \begin{tabular}{ccccccc} 
                \hline
  $\phi$     &$\tau_{edge}$ &N$_{H} $               &  Fe K$\alpha$                &apec normalisation   &nei  normalisation   &Fe K$\alpha$ EW\\
                     &  $10^{-2}$   & $10^{22}$cm$^{-2}$   &  $10^{-4}$ ph/s/cm$^2$    &$10^{-2}$ cm$^{-5}$            &$10^{-2}$ cm$^{-5}$            & keV     \\ 
                \hline
  2.9389    & $4.6 \pm 1.6$        & $3.09 \pm 0.17$            &  $4.59 \pm 0.23$            &  $11.70_{-0.63}^{+0.64}$            &  $10.08 \pm 0.31$    & 0.152   \\
  2.9393    & $3.9 \pm 1.1$        & $3.98 \pm 0.12$            &  $5.03 \pm 0.19$            &  $14.30_{-0.50}^{+0.51}$            &  $11.26 \pm 0.24$    & 0.144   \\
  2.9669    & $4.2 \pm 0.8$        & $3.82 \pm 0.08$            &  $8.92 \pm 0.19$            &  $18.21 \pm 0.49$                   &  $17.24 \pm 0.24$    & 0.180   \\
  2.9721    & $6.2_{-0.6}^{+1.1}$  & $5.62_{-0.10}^{+0.11}$     &  $8.87_{-0.23}^{+0.24}$     &  $< 0.29$                           &  $27.31 \pm 0.18$    & 0.182   \\
  2.9979    & $32.0_{-5.9}^{+6.0}$ & $9.58_{-1.79}^{+1.83}$     &  $0.88 \pm 0.04$            &  $0.25 \pm 0.15$                    &  $0.29 \pm 0.07$     & 1.102   \\
  3.0051    & $26.6_{-7.6}^{+7.8}$ & $48.41_{-10.03}^{+13.79}$  &  $1.50_{-0.12}^{+0.37}$     &  $0.05_{-0.05}^{+0.35}$             &  $0.30_{-0.12}^{+0.07}$  & 2.354  \\
  3.0089    & $16.0 \pm 4.3$       & $37.80_{-1.96}^{+1.99}$    &  $2.20 \pm 0.11$            &  $1.82_{-0.38}^{+0.40}$             &  $1.03 \pm 0.13$     & 0.567   \\
  3.0275    & $16.4 \pm 1.7$       & $13.09 \pm 0.24$           &  $3.66 \pm 0.11$            &  $8.56_{-0.28}^{+0.29}$             &  $1.88 \pm 0.12$     & 0.294   \\
  3.1719    & $16.3_{-3.2}^{+3.3}$ & $2.95 \pm 0.34$            &  $1.08 \pm 0.14$            &  $2.77 \pm 0.40$                    &  $3.38 \pm 0.21$     & 0.122   \\

                \hline
        \end{tabular}
\end{table*}

\subsection{High-energy power law}

\begin{figure}
  \centering
  \includegraphics[width=0.8\linewidth,clip]{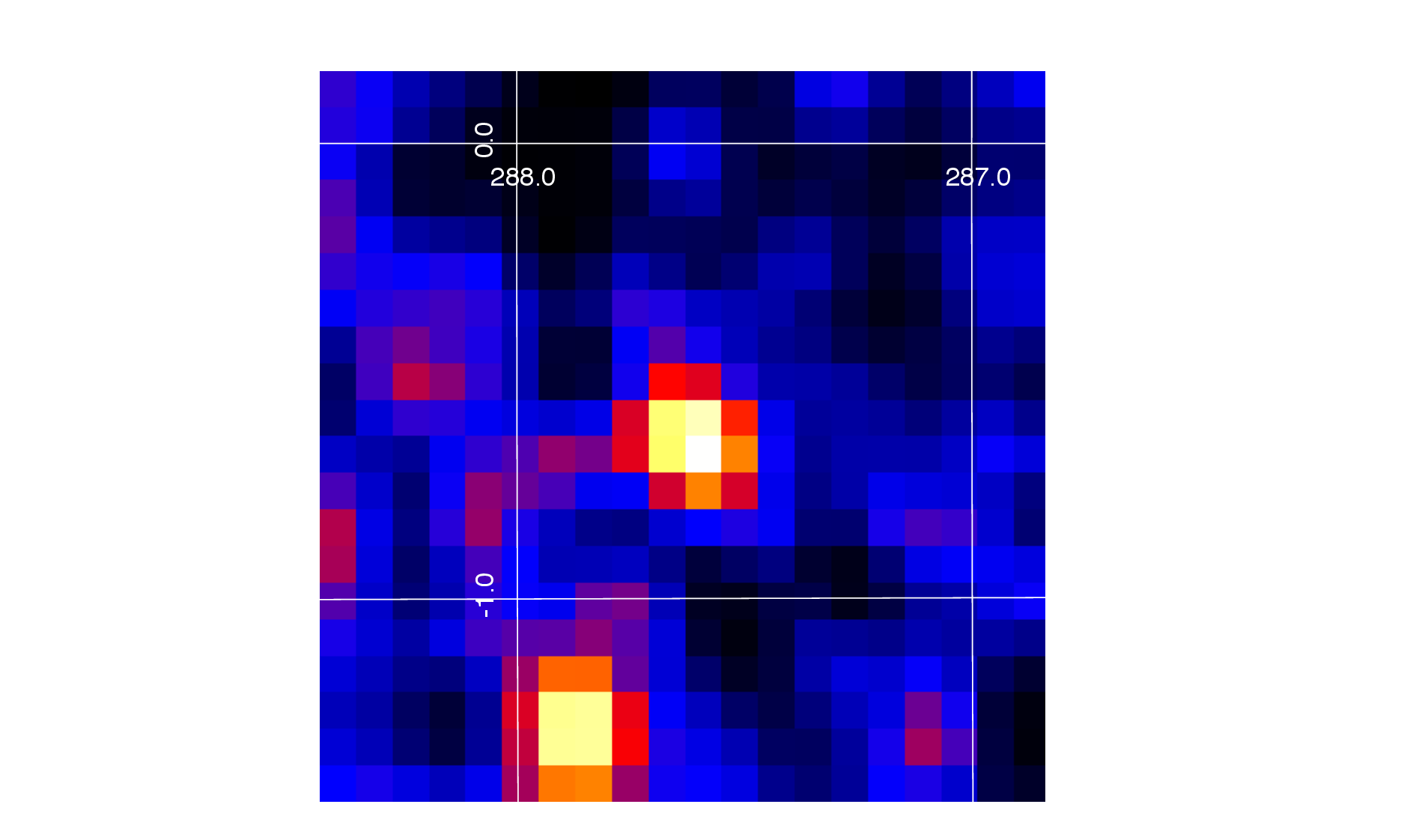}
  \caption[10]{ $\eta$ Carinae 40-80 keV image as observed by ISGRI on board INTEGRAL (Galactic coordinates).}
  \label{fig:img_integral}
\end{figure}

Previous works \citep{2010A&A...524A..59L, 2014ApJ...795..119H} have used Suzaku and INTEGRAL observations of $\eta$ Carinae to examine its emission at high energies. An INTEGRAL image of $\eta$ Carinae in the energy range 40-80 keV is shown in Fig. \ref{fig:img_integral}. It corresponds to observations conducted from January 11, 2013 to August 7, 2015 with a total effective exposure of $2.45\times10^6$ seconds. $\eta$ Carinae is located at the centre of this image and is clearly detected at these energies.

Both \cite{2010A&A...524A..59L} and \cite{2014ApJ...795..119H} found that the high-energy (above $\sim$20 keV) source flux can be described as a power law. However, our analysis of the NuSTAR data indicates that the power-law emission is faint, or even absent at energies lower than 30 keV. This was already found for two of the NuSTAR observations by \cite{2016ApJ...817...23H}.

\begin{figure}[]
  \centering
  \includegraphics[width=\linewidth, clip]{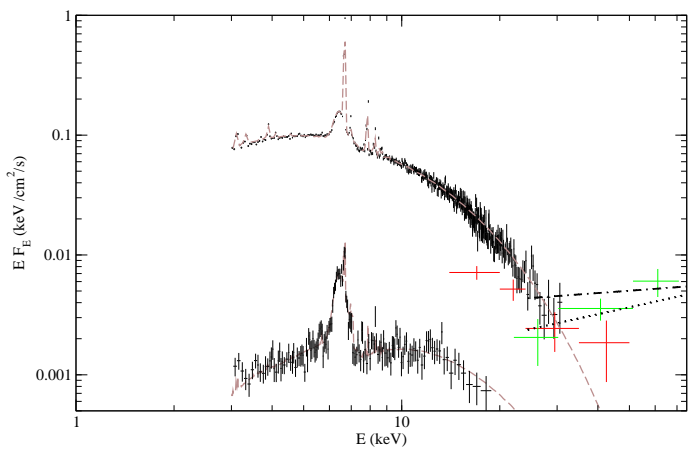}
  \caption[10]{X-ray spectra of $\eta$ Carinae. The black dots correspond to the spectra of the third and seventh NuSTAR observations and the red and green dots correspond to the BAT and INTEGRAL averaged spectra, respectively. The NuSTAR best-fit models are plotted as brown dashed lines. The dotted and dot-dashed lines correspond to the best-fit power law models to the Suzaku and INTEGRAL observations, respectively.}
  \label{fig:bat_isgri_spec}
\end{figure}

To investigate further this peculiar result, we produced the averaged source spectrum using the BAT and ISGRI data. Figure \ref{fig:bat_isgri_spec} compares the spectra measured by NuSTAR (black dots), BAT (red dots), and ISGRI (green dots). The NuSTAR spectra correspond to the brightest ($\phi=2.9669$) and faintest ($\phi=3.0051$) observation. The brown dashed lines denote the best-fit model of each spectrum, as described in the previous section. The Suzaku (dotted line) and INTEGRAL (dot-dashed line) best-fit power law models, as given in \cite{2014ApJ...795..119H} and \cite{2010A&A...524A..59L}, respectively, are also plotted for comparison. The Suzaku model corresponds to a spectral index of 1.4, while the index of INTEGRAL model is 1.8. 

We conclude that the power law is dominant above 30 keV, while the emission at lower energies can be explained as thermal WWC emission. The power law is too faint to be detected by NuSTAR with usual exposure time. Finally, if the power law extends below 30 keV, it should be variable.

\subsection{Variability of the WWC emission}

The X-ray flux (Fig. \ref{fig:lc_all}) varies similarly as observed during previous periastron passages \citep{2017ApJ...838...45C}. Hydrodynamic simulations \citep[e.g.][]{2011ApJ...726..105P} of the collision of the winds have succeeded in reproducing the overall variability of the source, where the long-term flux variability is attributed to variations in the structure (i.e. temperature and density) of the wind collision region and to variations of the column density of the absorption. However, the mechanism responsible for the $\sim$20 day-long minimum is still a matter of debate because it could be related to an increase of the absorbing material or the partial disappearance of the WWC if the ionisation state of the winds is too high or the distance between the stars becomes too small to allow the winds to reach high terminal velocities. Finally, the wind from the primary star may be powerful enough to stop the secondary wind.

In order to investigate which mechanism is consistent with the NuSTAR observations, we compared the observed source flux to the theoretically expected flux given in \cite{2017A&A...603A.111B}. We used our best-fit model to calculate the intrinsic 8-9 keV continuum emission of the WWC. The best-fit values of $N_H$, listed in Table \ref{tab:best-fit}, were used to correct the absorption. This correction only accounts for the extinction due to neutral atoms; the $N_H$ values were estimated using the \textit{wabs} model. To properly estimate the intrinsic WWC emission, we considered the obscuration by ionised particles (i.e. Thomson scattering) as well.

\begin{figure}
  \centering
  \vspace{0.52cm}
  \includegraphics[width=\linewidth,clip]{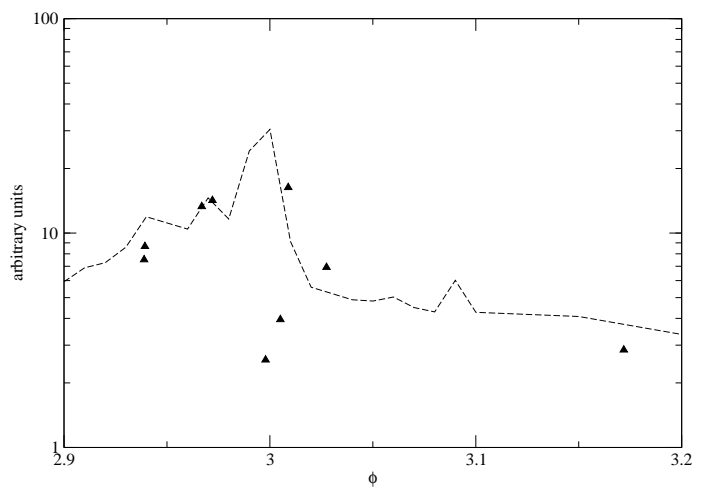}
  \caption[10]{Comparison between the observed continuum flux and flux deduced from simulation. The filled triangles indicate the intrinsic WWC flux corrected for both absorption and Thomson scattering. The dashed line denotes the expectation. The horizontal axis is the orbital phase.}
  \label{fig:lc_cont_vs_simul}
\end{figure}

A correction for Thomson scattering requires an estimation for the column density of the ionised particles. Since we cannot determine the column density directly, we used the emission-weighted column density of \cite{2011ApJ...726..105P} as our best guess. More precisely, we used the values they calculated for the model Orbit-IA (see Fig. 19 in their paper), which are listed in Table \ref{tab:nh-parkin}. 

The corrected fluxes are compared to the expected emission in Fig. \ref{fig:lc_cont_vs_simul}. The dashed line denotes the expected light curve and the filled triangles denote the observed WWC flux corrected for obscuration (scaled to arbitrary units). Except for the two observations around periastron, the corrected flux lies reasonably close to the predicted emission.  

The minimum, which cannot be explained by absorption effects, has been proposed to be related to the occultation of the WWC region. Although the inclination of  $\eta$ Carinae \citep[estimated as $130-145^{\circ}$;][]{2012MNRAS.420.2064M} renders the occultation of WWC by one of the stars improbable, we decided to study an extreme scenario. The time separation of the two observations in the minimum state is $\Delta \text{t}=14.57$ days (see the observations labelled 30002040004 and 30002010008 in Table \ref{tab:obs_info}). Assuming that our line of sight is on the orbital plane, that at periastron the primary star is closer to us than the secondary, and $\text{R}_\text{p} = 100 \hspace{2pt}\text{R}_\odot$ for the radius of the primary star, the size of the WWC region should be at least $3.5 \times \text{R}_\text{p}$ for the occultation to last  $\Delta \text{t}$. In this case the occultation would result in a maximum flux decrease of the order of $(\text{R}_\text{p}/{3.5\text{R}_\text{p}})^2 \simeq 8\%$. In Fig. \ref{fig:lc_cont_vs_simul} the decrease of the flux corresponds to a factor of 10; thus, the observed minimum cannot only be explained by occultation on. We conclude that the centre of the WWC zone partially vanishes during periastron.

\begin{table}
        \centering
        \caption{Column density of  ionised matter \citep{2011ApJ...726..105P} used to correct observed emission for Thomson scattering.}
        \label{tab:nh-parkin}
        \begin{tabular}{cc} 
                \hline
                \noalign{\smallskip}
  $\phi$   &   N$_{H} (10^{23}\text{cm}^{-2})$   \\
  \noalign{\smallskip}                       
\hline 
\noalign{\smallskip}
  2.9389    & 1.5  \\
  2.9393    & 1.5  \\           
  2.9669    & 2.5  \\             
  2.9721    & 2.5  \\ 
  2.9979    & 40   \\ 
  3.0051    & 50   \\
  3.0089    & 45   \\ 
  3.0275    & 15   \\        
  3.1719    & 5    \\      
  \noalign{\smallskip}
\hline 
        \end{tabular}
\end{table}

The X-ray minimum occurs right at periastron, about 40 days before the local $\gamma$-ray minimum detected by \cite{2017A&A...603A.111B}. These two minima are not related. The X-ray minimum is short and corresponds to the likely partial disappearance of the WWC. The gamma-ray minimum corresponds to the change of the shock front geometry with the appearance of a reverse shock bubble developing after periastron.

\subsection{Fe K$\alpha$ line emission}
\label{sec:Fe_line}

The width of the Fe K$\alpha$ line ($\sigma = 0.22$ keV, i.e. $10^4$ km/s) is too large to be explained by bulk motions or thermal broadening of the iron nuclei. Such broad lines have been observed in low-mass X-ray binaries and interpreted \citep{1989ApJ...341..955K} with Compton broadening where the Fe K$\alpha$ photons are scattered by a population of thermal electrons. The observed broadening, estimated as $\sigma \sim 0.12\cdot \tau\cdot (1+0.78\cdot \texttt{kT}/1~\texttt{keV})$, requires electrons reaching energies $\sim$1 keV, similar to those of the thermal emission generated in the shock of the WWC region. It is therefore tempting to assume that iron fluorescence largely takes place in the dense pre-shock zone of the outer parts of the WWC region, surrounded by the shock fronts, where Compton scattering occurs. 

The variability of the broad iron line flux (see lower panel of Fig. \ref{fig:best_fit_1}) can be used to estimate the size of its emitting region. The line flux varies significantly (above a 10$\sigma$ level) in 22 days between the two observations at $\phi = 2.9979$ and $\phi = 3.0098$. As a result, the size of the iron line emitting region is lower than 22 light days and is probably much lower if the iron nuclei are indeed broadened by the hot electrons. The faint broad Fe line observed by \cite{2004ApJ...613..381C} in the HN probably results from the scattering of the broad Fe line photons that we detected. 

\begin{figure}
  \centering
  \vspace{0.65cm}
  \includegraphics[width=\linewidth,clip]{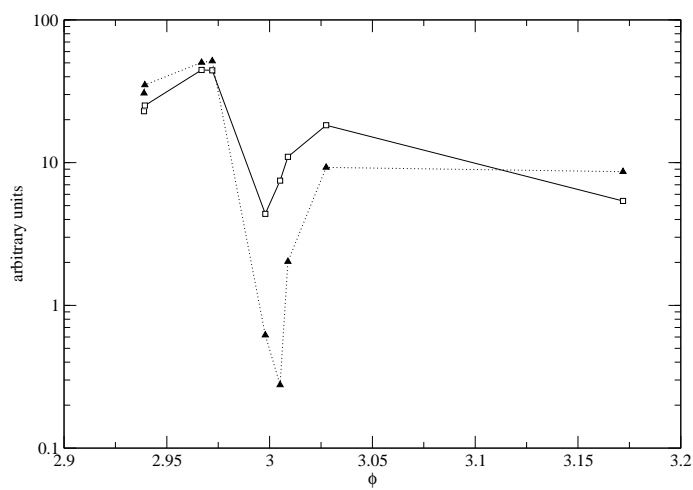}
  \caption[10]{Comparison between observed WWC flux (filled triangles) and Fe line normalisation (open squares) scaled to allow a direct comparison. The horizontal axis is the orbital phase.}
  \label{fig:lc_cont_vs_fe}
\end{figure}

The line variability appears to be similar to that of the X-ray continuum. This is clearly shown in Fig. \ref{fig:lc_cont_vs_fe} where the open squares denote the best-fit iron line normalisations and the filled triangles indicate the observed (without any absorption correction) WWC X-ray fluxes. The similar variations indicate that the WWC is the main source of the Fe fluorescence line seed photons.

\begin{figure}
  \centering
  \vspace{0.65cm}
  \includegraphics[width=\linewidth, clip]{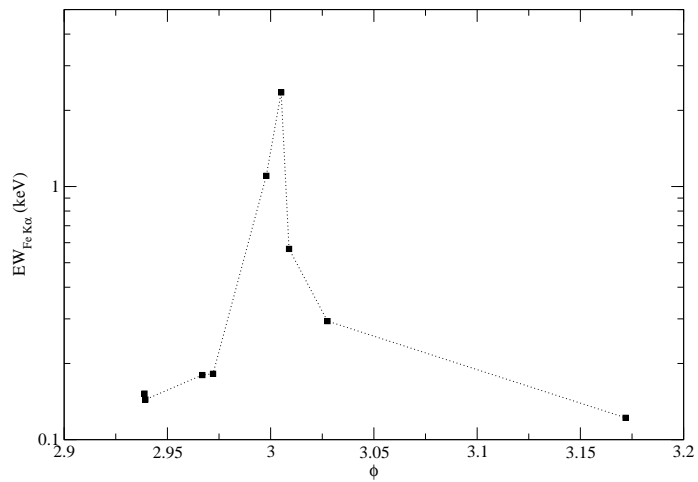}
  \caption[10]{Equivalent width of the Fe line. For the calculation, we used the observed continuum without any correction for absorption.}
  \label{fig:fe_ew}
\end{figure}

The variability amplitude of the Fe line is however significantly smaller than that of the X-ray continuum. To investigate the origin of this difference, we calculated the equivalent widths of the broad Fe line, which are listed in Table \ref{tab:best-fit} and  plotted in Fig. \ref{fig:fe_ew}. The large equivalent width observed for the three observations around periastron indicates a strong line emission during this period, despite the apparent decrease of the continuum. This could be understood if the obscuration of the brightest part of the WWC is stronger than that affecting the likely more extended Fe line emitting region.

\section{Conclusions}

We analysed 10 NuSTAR observations of $\eta$ Carinae to investigate its X-ray spectral variability around the 2014 periastron. After extracting high-level products for each observation, we modelled the spectra with a model consisting of a constant component (fitted to the X-ray minimum at periastron) and a variable component made of equilibrium and non-equilibrium thermal emission and of broad Fe K$\alpha$ and $\beta$ emission lines. Our main results can be summarised as follows.

The high-energy power-law emission, detected in $\eta$ Carinae by INTEGRAL and Suzaku, is too faint below 30 keV to be detected by NuSTAR, as already discussed by \cite{2016ApJ...817...23H}. The normalisation of that component is certainly variable unless it would feature an unexpectedly sharp low energy cut-off. The sensitivity of current instruments does not allow monitoring of the variability of the hard X-ray spectral component on timescales relevant to the periastron passage.
                  
For most observations, the variable continuum component matches reasonably well the observations obtained during previous periastrons and the expected emission of the central region of the WWC zone, when corrected for absorption and scattering. The flux observed within $\sim 10$ days of the periastron is too low when compared to the expectations and cannot be explained solely by absorption or occultation effects. This suggests a partial disruption of the central region of the WWC zone at periastron, for example related to a temporary interruption of wind acceleration for one of the two stars. 
            
A broad Fe K$\alpha$ emission line is detected at 6.4 keV; the flux and equivalent width of this emission line are variable. The line broadening likely occurs because of Compton scattering by the electrons heated in wind collision shock. Its equivalent width is a complex function of the fluorescence region geometry and of its relative obscuration when compared to the main continuum emission region.
                  
\begin{acknowledgements}
We thank the NuSTAR team and principal investigator for the scheduling of the $\eta$ Carinae campaign.
\end{acknowledgements}

\bibliographystyle{aa} 
\bibliography{draft_arxiv} 

\end{document}